\documentclass{aa}  

\usepackage{soul}
\usepackage{graphicx}
\usepackage{color}
\usepackage{txfonts}
\usepackage{natbib}
\usepackage[squaren,Gray]{SIunits}

\newcommand{\gcc}{\mbox{ g cm$^{-3}$}}
\usepackage{bbold}
\usepackage{subcaption}
\usepackage{mwe}

\begin{document}
  \renewcommand{\vec}[1]{\mathbf{#1}}

   \title{Dust dynamics in current sheets within protoplanetary disks }
\subtitle{I. Isothermal models including ambipolar diffusion and Ohmic resistivity.}
   \author{U. Lebreuilly
          \inst{1,2,3}
          \and M.-M. Mac Low\inst{3} \and
          B. Commer\c con\inst{2} \and D. S. Ebel\inst{4}  }

   \institute{Université Paris-Saclay, Université Paris Cité, CEA, CNRS, AIM, 91191, Gif-sur-Yvette, France  \and \'Ecole normale sup\'erieure de Lyon, CRAL, UMR CNRS 5574, Universit\'e de Lyon , 46 All\'ee d Italie, 69364 Lyon Cedex 07, France \and Department of Astrophysics, American Museum of Natural History, Central Park West at 79th Street, New York, NY 10024, USA \and Department of Earth and Planetary Sciences, American Museum of Natural History, Central Park West at 79th Street, New York, NY 10024, USA 
              \\
              \email{ugo.lebreuilly@cea.fr}      
             }

   \date{}

 
  \abstract
   {Chondrules originate from the reprocessing of 
   dust grains. They are key building blocks of telluric planets, yet their formation, which must happen in strongly localized regions of high temperature, remains poorly understood. }
   {We examine the dust spatial distribution near regions of strong local heating produced by current sheets, as a step toward exploring a potential path for chondrule formation. We further aim to investigate current sheet formation under various conditions in protoplanetary disks in the presence of ambipolar diffusion and Ohmic resistivity and the effect of current sheet morphology on dust dynamics in their vicinity.}
   {We use the \texttt{RAMSES} code including modules for non-ideal magnetohydrodynamics and solution of the dynamics of multiple sizes of dust grains to compute unstratified shearing box simulations of current sheet formation. We investigate, through seven models the effect of the ambipolar diffusion and Ohmic resistivity strength, the initial density, and magnetic field, as well as the resolution and box size.}
   {We find that current sheets form in all our models, with typical widths of $10^{-3}$--$10^{-2}$~AU and that strong dust fraction variations occur for millimeter-sized grains. These variations are typically of an order of magnitude and up to two orders of magnitude for the most favorable cases. We also show that the box size and resolution has a strong impact on the current sheet distribution and intensity. }
   {The formation of current sheets that can intensely heat their surroundings near strong dynamical dust fraction variations could have important implications for chondrule formation, as it
appears likely to happen in regions of large dust fraction.}

   \keywords{Hydrodynamics; Magnetohydrodynamics (MHD); Turbulence;  Protoplanetary disks;  Interplanetary medium; Planets and satellites: formation }

  \maketitle

\section{Introduction}

The rock content of the Solar System originates primarily from interstellar dust.  The earliest rocks formed in the Solar System, chondrites, were highly processed during early disk evolution.  Chondrites
 are the most often found type of meteorites (around 80\% by mass). In these meteorites, the chondrules represent 20--80\% of the mass. They are surrounded by a matrix of fine grains that are more numerous than chondrules. The chondrules are spherical silicate grains of 0.1--1~$\milli\meter$ with a glassy texture \citep{2005ASPC..341..251J}. To constrain the physical conditions at play in protoplanetary disks, it is essential to find a mechanism common enough to explain chondrule formation that can reproduce their apparent abundance in the Solar System. 

Chondrule textures indicate that they form during rapid heating and cooling events called flashes. These must meet at least four prerequisites: they must have extremely short heating timescales \citep[less than a few minutes,][]{1998Sci...280...62C}; be very localized so that the chondrules can exit them rapidly \citep{2015Icar..245...32H}; be energetic enough to increase the grain temperature up to $\sim 1700$--$2000~\kelvin$ \citep{1990GeCoA..54.3537L,1990GeCoA..54.3475R,1996cpd..conf..197H}; and, compared to the free-space cooling time of chondrule-sized objects of a few seconds, the cooling rate in and near these flashes must be relatively slow \citep[$\approx 10^2-10^3~\kelvin ~\hour^{-1}$][]{1990GeCoA..54.3475R}. 
The abundance of chondrules is evidence that these flashes, no matter how and where they occur, are common enough to turn a large fraction of the dust into chondrules. 

A significant fraction of chondrules have multiple rims that must have
formed in repeated flashes during their formation \citep{2020E&PSL.54216286B}.  Chondrule formation theories must also explain their narrow range of sizes \cite[0.1--1~$\milli\meter$,][]{2014Icar..232..176J,2015ChEG...75..419F}, their diversity of compositions, and the presence of volatile-rich matrix \citep{2018crpd.book..151E}. The evidence for complementarity suggests a reservoir of common origin for chondrules and matrix \citep{2015E&PSL.411...11P,2005PNAS..10213755B}. Nevertheless, the matrix grains are clearly different from chondrules in term of composition and size. They contain a substantial abundance of volatiles that would evaporate at temperatures higher than 500--800 $\kelvin$,  which indicates that they do not experience any dramatic heating events. In addition, the matrix is  mostly composed of fine $\lesssim 5 \micro\meter$ grains. Whether it is a by-product of chondrule formation \citep{2005ASPC..341..701H} or a population that experiences a totally different evolution, the matrix and chondrule formation are intrinsically related \citep{2018crpd.book..151E}. Last but not least, high dust concentrations seem required to explain the volatile abundances \citep{2008Sci...320.1617A} in chondrules, which suggests that their precursors should be able to concentrate preferentially while the matrix would stay well coupled to the gas.

Magnetic fields in protoplanetary disks have been widely studied as a possible source of angular momentum transport via the magnetorotational instability \citep[hereafter MRI,][ among others]{1991ApJ...376..214B,1996ApJ...463..656S,2000ApJ...543..486S}.  As disks are in fact poorly ionised, the MRI could be inhibited in disk midplanes in the so-called dead zones \citep{1996ApJ...457..355G,2000ApJ...530..464F}.
More recent studies have suggested that it might even be suppressed entirely in disks, with the angular momentum carried away instead by magnetocentrifugal winds  \citep{2013ApJ...769...76B,2014A&A...566A..56L}. An imperfect coupling between the neutral and the magnetic field, however, might give birth to dissipative structures  \citep{1994ApJ...427L..91B}.

An interesting hypothesis for chondrule formation is that it occurs in or around thin
current sheets \citep{2004ApJ...606..532J}. These dissipative structures are believed to occur because of finite electric conductivity \citep{1972ApJ...174..499P,1994ISAA....1.....P,2004ApJ...606..532J}. The resistive heating might be efficient enough in current sheets so that dust grains reach their melting temperatures \citep{2014ApJ...791...62M}.  In addition, it was shown by \cite{2012ApJ...761...58H} that the increase of the temperature could produce resistivity gradients that make the sheets thinner, making them a potentially favoured place for chondrule formation.

 Recent studies have started to investigate dust dynamics in resistive disks \citep{2018A&A...617A.117R,2020arXiv200601194R}. However the dust behaviour at the smaller scales of current sheets remains essentially unexplored. In addition, current sheet formation with Ohmic resistivity has been investigated in the unstratified shearing box models of \cite{2014ApJ...791...62M} and more recently by \cite{2018MNRAS.477.3329R} but the effect of ambipolar diffusion on current sheet formation has not been investigated. In this work (paper I), we aim to understand the dynamical sorting of dust grains in the vicinity of current sheets. Moreover, we want to investigate current sheet formation in the presence of both ambipolar diffusion and Ohmic dissipation. With that in mind, we perform simulations of dissipative current sheet formation using the shearing box \citep{2018A&A...620A..21C} and dust dynamics modules \citep{2019A&A...626A..96L} of the {\ttfamily RAMSES} code \citep{2002A&A...385..337T,2006A&A...457..371F}.
 
 
This manuscript is organised as follows. In Sect.~\ref{sec:theory}, we detail the theoretical framework used in this study. After that, we explain our method of solution in Sect.~\ref{sec:methods}. Then, in Sect.~\ref{sec:models}, we introduce and describe the different models that we computed. Following this, we discuss our results in Sect.~\ref{sec:discuss} and finally present our conclusion and prospects in Sect.~\ref{sec:conclu}.

  \section{Theoretical framework}
  \label{sec:theory}
  \subsection{Gas and dust coupling}

  We describe gas and dust mixtures with a monofluid approach in the terminal velocity approximation \citep[see][for more details on gas and dust monofluids]{2014MNRAS.440.2136L,2014MNRAS.440.2147L,2014MNRAS.444.1940L,2018MNRAS.476.2186H,2019A&A...626A..96L,2020A&A...641A.112L}. The mixture of density $\rho$ is composed of a plasma (neutral atoms, ions, and electrons) and neutral dust grains. The plasma has a density $\rho$ and velocity $\vec{v}$ and the dust fluids each have a density $\rho_k$ and velocity $\vec{v_k} \equiv
  \vec{v} +\vec{w_k}$, where $\vec{w_k}$ is the differential velocity between the dust and the plasma.
\subsection{Dusty non-ideal MHD for neutral grains} 

Around T-Tauri stars, the disk mass is much smaller than the mass of the star,
so one can neglect the self gravity of the disk. In this context, the equations of magnetohydrodynamics (MHD) with $\mathcal{N}$ neutral dust species denoted by index $k$ can be written as 
\begin{eqnarray} 
\label{eq:disk}
 \frac{\partial \rho}{\partial t}  + \nabla \cdot \left[ \rho \vec{v} \right]&=&0, \nonumber  \\   \frac{\partial   \rho_{k} }{\partial t} +\nabla \cdot \left[ \rho_{k} \left( \vec{v} +  \vec{w_k}\right) \right] &=& 0, \ \forall k \in \left[1,\mathcal{N}\right], \nonumber \\
  \frac{\partial \rho \vec{v}}{\partial t} + \nabla \cdot \left[\left(P_{\mathrm{g}}+\frac{\vec{B}^2}{2}\right) \mathbb{I} + \rho (\vec{v}\otimes  \vec{v}) -\vec{B}\otimes \vec{B} \right] &=&-\rho \vec{g}, \nonumber \\
    \frac{\partial \vec{B}}{\partial t} - \nabla \times \left[\vec{v}\times \vec{B}\right] & =&-\nabla \times \vec{E_p},\nonumber \\
    \nabla \cdot  \vec{B} &=&0, 
 \end{eqnarray}
 where $P_{\mathrm{g}}$ is the gas pressure, $\vec{B}$ is the magnetic field, $\vec{g}$ is the gravitational acceleration, and where the dust differential velocity for each species $k$ is given by \citep{2020A&A...641A.112L}
\begin{eqnarray}
 \vec{w_k} \equiv \left[\frac{\rho}{\rho-\rho_k}t_{\mathrm{s},k}- \sum_{l=1}^{\mathcal{N}} \frac{\rho_l}{\rho-\rho_l} t_{\mathrm{s},l}\right] \frac{\nabla P_{\mathrm{g}} - \vec{J}\times \vec{B}}{\rho},
  \end{eqnarray}
where $\vec{J}\equiv\nabla \times \vec{B}$ is the electric current.
  The stopping time \citep{1924PhRv...23..710E}
  \begin{eqnarray}
  t_{\mathrm{s},k} \equiv \sqrt{\frac{\pi \gamma}{8}} \frac{\rho_{\mathrm{grain},k}}{\rho}\frac{s_{\mathrm{grain},k}}{c_{\mathrm{s}}},
  \end{eqnarray}
  where grain species $k$ has radius $s_{\mathrm{grain},k}$.
Neglecting the Hall effect because it does not dissipate energy, as $\left((\nabla \times \vec{B})\times \vec{B}\right) \cdot (\nabla \times \vec{B}) = 0$, the electric field in the comoving frame of the plasma (i.e. everything except the dust)
  \begin{eqnarray}
  \label{eq:ohmlaw}
  \vec{E}_{\mathrm{p}}\equiv  - \eta_{\mathrm{O}} (\nabla \times \vec{B})+ \frac{\eta_{A}}{|\vec{B}|^2}((\nabla \times \vec{B}) \times \vec{B}) \times \vec{B},
  \end{eqnarray}
where $\eta_{\mathrm{O}}$ and $\eta_{\mathrm{A}}$ are the Ohmic and ambipolar resistivities. At this point, let us also recall the definition of the plasma parameter  \begin{eqnarray}
 \beta \equiv \frac{2 P_{\mathrm{g}} }{|\vec{B}|^2}.
 \end{eqnarray}
 
\subsection{Unstratified shearing box approximation}

Modeling the protoplanetary disk as a whole is computationally expensive, especially when attempting to resolve thin current sheets. Fortunately, a simple approximation can be made when considering only a small part of the disk. In the shearing box approximation \citep{1995ApJ...440..742H}, we only model a small volume of the disk at a radius $R_0$ with a length $L_0\ll R_0$ typically of a few scale heights $H$ and in rotation at the Keplerian velocity $\Omega_{\mathrm{kep}} (R_0)$\footnote{for simplicity we  write $\Omega \equiv \Omega_{\mathrm{kep}} (R_0)$}. In this context and neglecting the vertical stratification of the disks, the total momentum conservation equation becomes \citep{1995ApJ...440..742H}
\begin{eqnarray}
\label{eq:diskshear}
\frac{\partial \rho \vec{v}}{\partial t} + \nabla \cdot \left[\rho \vec{v} \otimes \vec{v} + \left(P_{\mathrm{g}} +\frac{\vec{B^2}}{2}\right) \mathbb{I} + \vec{B}\otimes \vec{B}\right]= -2  \rho \vec{\Omega} \times \vec{v} + \rho \vec{g},
\end{eqnarray}
where $\vec{g} = -2 q \Omega^2 \vec{x} $, and $\vec{x}$ is the position along the radial axis. The parameter $q$ depends on the radial profile of the disk angular velocity and is equal to 3/2 in the Keplerian case. The term $-2 \rho q \Omega^2 \vec{x}$ represents the centrifugal pseudo-force and $ -2  \rho \vec{\Omega} \times \vec{v}$ is the Coriolis pseudo-force.

\subsection{Elsasser numbers}
\label{sec:resi}

At this stage it is useful to define the Elsasser numbers $\rm{Am}$ and $\Lambda$ of the ambipolar and Ohmic diffusion, respectively\footnote{or $\mathrm{Els}$ when the source of non-ideal MHD is not specified}. They quantify the relative importance of resistive effects and Alfv\'en wave propagation
and are defined as
\begin{eqnarray}
\rm{Am} &\equiv& \frac{v_{\rm{A}}^2}{\eta_{\rm{A}}\Omega},\nonumber \\
\Lambda &\equiv&\frac{v_{\rm{A}}^2}{\eta_{\rm{O}}\Omega},
\end{eqnarray}
where $v_{\rm{A}}  \equiv {|\vec{B}|}/{\sqrt{\rho}}$ is the Alfv\'en speed. In this work, we use these numbers to impose the initial values of the resistivities. According to these definitions of the Elsasser numbers the corresponding resistivity (either $\eta_{\rm{A}}$ or $\eta_{\rm{O}}$) is
\begin{eqnarray}
\eta_{\rm X} = 3 \times 10^{14} \left(\frac{\mathrm{Els}}{1}\right)^{-1} \left(\frac{\beta}{750}\right)^{-1}  \centi\meter^2 \second^{-1}.
\end{eqnarray}
We consider two values for the Elsasser numbers to model strong ($\mathrm{Els}=1$) and weak ($\mathrm{Els}=10$) resistivity cases. These values are typical for protoplanetary disks interiors \citep[see the recent review by][and references therein]{2021JPlPh..87a2001P}.

Physically, the resistivity is determined by the complex interplay between ionization and recombination on dust grains and in the gas \citep[e.g.][]{2015ApJ...811..156D}. However, by setting fixed Elsasser numbers, we can examine how current sheets behave in the regime where resistivity is important enough to produce heating but not so dominant that it suppresses the formation of current sheets.

\subsection{Resistive heating}

The Ohmic and ambipolar resistivities each introduce a heating term in the energy equation
\begin{eqnarray}
\label{eq:ohmicheat}
\Gamma_{\mathrm{O}} &\equiv &\eta_{\rm{O}} ||\vec{J}||^2,\nonumber \\
\Gamma_{\mathrm{A}} &\equiv &\eta_{\rm{A}}\frac{ ||\vec{J}\times\vec{B}||^2}{||\vec{B}||^2}.
 \end{eqnarray}

 In this preliminary study, we use the isothermal approximation for the gas, so these terms are
 not included in the calculation but instead estimated in post-processing.

 \subsection{Magnetorotational instability}
 In magnetised disks, the interplay between the differential rotation and the tension of the magnetic field lines can lead to the MRI \citep{1991ApJ...376..214B}. In the ideal MHD case, where the coupling between the fluid and the magnetic field is perfect, one finds the wavelength for the fastest growing mode in terms of the scale height \citep{1995ApJ...440..742H,2011ApJ...736..144B} 
  \begin{equation}
     \frac{\lambda_{\rm{c}}}{H} = \frac{9.18}{\sqrt{\beta}}.
 \end{equation}
 When the resistive effects are important, the expression for the wavelength of the fastest growing mode is modified. In the ambipolar case it is \citep{1999MNRAS.307..849W,2011ApJ...736..144B}
 \begin{equation}
     \frac{\lambda_{\rm{c}}}{H} = \frac{5.13}{\sqrt{\beta}}\sqrt{1+\frac{1}{\rm{Am}^2}}.
 \end{equation}
This rough estimate of $\frac{\lambda_{\rm{c}}}{H}$ allows us to determine if the simulation box that we use is comfortably larger than the fastest growing mode, as is necessary to properly capture the MRI.
 \subsection{Angular momentum transport}
The usual way to estimate the transport of angular momentum in protoplanetary disks is to compute the viscosity parameter \citep{1976MNRAS.175..613S} 
\begin{eqnarray}
\alpha = \frac{\left< \rho u_{r}u_{\phi} - B_r B_{\mathrm{\phi}}\right>}{\left<P_{\mathrm{g}}\right>}
\end{eqnarray}
where $u_r$ and $u_{\mathrm{\phi}}$ are the radial and azimuthal components of the gas velocity relative to the shear. Typically, $\alpha<1$ in the case of resistive MRI and can be much smaller if the instability is damped. To compare the models, we measure $\bar{\alpha}$, which is the time averaged $\alpha$ over the last $10~\Omega^{-1}$ of the run to make sure that the saturated regime of the turbulence is reached and that the measure is not affected by the initial conditions.

\subsection{Current sheet analysis}
\label{sec:csheetanalys}
Following \citet{2018MNRAS.477.3329R}, we define  Ohmic and ambipolar current sheets as regions of strong dissipation. We compute the following quantities 

\begin{eqnarray}
\varepsilon_{\mathrm{O}}&=& \left<\Gamma_{\mathrm{O}}\right> + 3 \sigma_{\Gamma_{\mathrm{O}}}, \nonumber \\
\varepsilon_{\mathrm{A}}&=& \left<\Gamma_{\mathrm{A}}\right> + 3 \sigma_{\Gamma_{\mathrm{A}}},
\end{eqnarray}
 In the rest of this work, and similarly to \cite{2018MNRAS.477.3329R}, we define a region of strong dissipation due to a non-ideal effect $\theta$ (with $\theta = $~O for Ohmic or A for ambipolar) as
\begin{equation}
\label{eq:defcs}
\Gamma_{\theta}> \varepsilon_{\theta}.
\end{equation}
Those regions thus have dissipation rates more than three standard deviations above the average value.

\section{Numerical Methods}
\label{sec:methods}
\subsection{Numerical scheme}

In this work, we use the \texttt{RAMSES} code \citep{2002A&A...385..337T,2006A&A...457..371F} and its dust dynamics solver \citep{2019A&A...626A..96L} extended to neutral grains in MHD by \cite{2020A&A...641A.112L}. We also use the implementation of ambipolar diffusion and Ohmic resistivity of \citet{2012ApJS..201...24M}.

We integrate Eqs.~\ref{eq:disk} replacing the momentum equation by Eq.~\ref{eq:diskshear}.  We use the MUSCL scheme of \texttt{RAMSES} with the HLLD Riemann solver for the barycenter part of the MHD equations and for the induction equation \citep{2005JCoPh.208..315M}. For stability and similarly to \cite{2013A&A...552A..71F}, the solver automatically switches to a Lax–Friedrichs solver where $\beta<10^{-3}$. As in  \cite{2013A&A...552A..71F} and again for stability reasons, we use the multidimensional slope limiter of \cite{10.1137/S1064827599360443} for the barycenter part of the conservation equations.  Similarly to  \cite{2018A&A...620A..21C}, we use an operator-splitting and an implicit Crank-Nicholson scheme to take into account the shear source terms in Eq.~\ref{eq:diskshear} without adding any constraint on the stability of the scheme. For the dust differential advection term, we use the dust solver of \cite{2019A&A...626A..96L,2020A&A...641A.112L} with the minmod slope limiter. 

When regions of very small density form in a model, they can lead to very large Alfv\'en velocities and hence very small timesteps. When this happens, evolving the model  significantly in a reasonable computational time becomes impossible. To circumvent this issue, we  impose an adaptive density floor that prevents $\beta$ from dropping below the value $\beta_{\rm{min}}= 10^{-4}$. This requires that the density be
\begin{equation}
\rho = \rm{max} \left(\rho, \beta_{\rm{min}}\frac{|\vec{B}|^2}{2 c_{\rm{s}}^2}\right).
\end{equation} 
 We point out that this method is strictly equivalent to imposing a maximum Alfv\'enic Mach number. We verified that the total box mass is not much affected throughout the calculation as it is conserved within about $1\%$ in all models. 

 We impose a maximum dust differential velocity of $5~ \kilo\meter ~\second^{-1}$ everywhere in the box to avoid unrealistically large dust velocities or new constraints on the timestep. This value is safely higher than the gas sound-speed which is around $1~ \kilo\meter ~\second^{-1}$  and is typically only reached in regions of very low densities. In such low density regions, the terminal velocity approximation no longer holds. For safety, we also enforce the maximum Stokes number to be $0.3$  by setting 
 \begin{eqnarray}
{t_{\rm{s},k}} = \rm{min} \left(t_{\rm{s},k}, \frac{0.3}{ \Omega} \right).
\end{eqnarray}  
This is similar to the method of \cite{2018MNRAS.477.2766B}, but we use the Stokes number while their regularisation was based on the stability condition of their scheme.

\subsection{Setup}

We impose a uniform initial density ($10^{-11}$ g cm$^{-3}$) and a uniform initial temperature of $300~\kelvin$. The initial magnetic field is vertical with $\beta_0=750$, except for the $\textsc{O10A10beta7500}$ run, which is initialized with $\beta_0=7500$. Unless specified, these models have a box size of one scale height, i.e $0.05~$AU, as all the models are computed at $R_0=1~$AU and we assume $H/R=0.05$. All the models are computed in the isothermal and unstratified shearing box approximations. 

In all the runs, we consider three dust sizes of $10~\micro\meter$ (St = $4 \times 10^{-4}$), $100~\micro\meter$ (St = $4 \times 10^{-3}$), and $1~\milli\meter$ (St =$4 \times 10^{-2}$), with initial dust ratios of 1/300. This leads to a total dust-to-gas ratio of 1/100. We choose not to explore the behavior of grains smaller than a micron because they would be strongly coupled with the gas (similarly to what we already observe for the $10~\micro\meter$ grains).  In addition, we do not study larger grains because the largest chondrules are smaller than a few millimeters \citep{2015ChEG...75..419F}. All the dust grains have an intrinsic grain density of $3~\gcc$ which is in line with the typical density of chondrules \citep[e.g.][]{1980E&PSL..51...26H,2014Icar..232..176J,2015ChEG...75..419F}. Note that initial dust concentration is unimportant as long as the dust back-reaction is weak, so we present the relative dust fraction variations $\bar{\epsilon} =\epsilon/\epsilon_0$  in this study rather than the actual dust fractions $\epsilon$ of the models.

 The  initial ambipolar and Ohmic resistivities are uniform and imposed by setting the value of the Elsasser numbers. Throughout the run, the Ohmic resistivity stays constant.   The ambipolar resistivity however varies, as it scales $\propto |\vec{B}^2|$ (see Eq.~\ref{eq:ohmlaw}). To avoid very small timesteps, we cap the value of the resistivities by the value $10\Omega H^2$ as was done by \cite{2014A&A...566A..56L}.

In all models, azimuthal and vertical boundaries are treated as simple periodic boundaries. The radial boundaries are, however, treated according to the shearing box implementation of \cite{2018A&A...620A..21C} that we adapted to Keplerian rotation by setting $q=3/2$.

\section{Results}
\label{sec:models}


\begin{table*}
  \caption{Summary of the different simulations. All of them are computed for $50\Omega^{-1}$, except for $\textsc{O10A10-HRES}$, which is only computed $26\Omega^{-1}$ for numerical cost reasons and $\textsc{O10A10beta7500}$, which is computed for $200\Omega^{-1}$, as it takes a longer time to reach a steady state. Res. stands for resolution. $\bar{\alpha}$ is averaged over $10~\Omega^{-1}$ at the end of the simulation. Dimensional quantities are given in cgs units. Note that the models all have the same Stokes numbers for the 3 grain sizes St = ($4 \times 10^{-4}$, $4 \times 10^{-3}$, $4 \times 10^{-2}$).}      
\label{tab:modelsdisk}      
\centering          
\begin{tabular}{c c c c c c c c c c c c c c }     
\hline\hline       
                   Model &  Res. & $L_{\rm{box}}$ [H] &  $\beta$ & $\rm{Am}$ & $\Lambda$  &  $\lambda_{\rm{c,th}}/H$  & $\bar{\alpha}$ & $\left<\Gamma_O\right>$& $\left<\Gamma_A\right>$ & $\sigma_{\Gamma_O}$& $\sigma_{\Gamma_A}$  \\ 
\hline                    
$\textsc{O10A10}$&   $128^3$ & 1  &    $750$ &  10 & 10 &$0.18$  & 1.3(-1) &  $4.8 (-9)$ & $8.7(-7)$& $ 1.2(-8)$ &$2.8(-6)$  \\
$\textsc{O10A10-HR}$&   $256^3$ & 1 &   $750$ &  10 & 10  &$0.18$ & 9(-2)  & $3.5(-9)$ & $2.9(-7)$& $ 7(-9)$ &$7.3(-7)$ \\
$\textsc{O10A10-LARGE}$&   $256^3$ & 2 &   $750$ &  10 & 10  & $0.18$ & 3.7(-2) &$1.5(-9)$ & $1.2(-7)$& $ 3.5(-9)$ &$3.1(-7)$  \\
$\textsc{O10A10beta7500}$&   $128^3$ & 1  &   $7500$ &  10 & 10&$0.1$  & 3(-3)& $5.4(-11)$ & $4.4(-8)$& $ 7.9(-11)$ &$1.0(-8)$ \\
$\textsc{O1A1}$&   $128^3$ & 1 &  $750$ &  1 & 1 & $0.26$& 2.5(-2)   & $3.2(-9)$ & $8.6 (-8)$& $ 4.8(-9)$ &$1.6(-7)$   \\
$\textsc{O10A1}$&   $128^3$ & 1 &  $750$ &  1 & 10  &$0.26$ & 1(-1) &  2.5(-9) & 2.8(-6)& 5.4(-9) &7.7(-6)   \\
$\textsc{O1A10}$&   $128^3$ & 1 &  $750$ &  10 & 1&$0.18$ & 1.6(-2)    & 2.1(-9) & 3.2(-9)& 2.7(-9) &6.4(-9)   \\

\hline \hline
\end{tabular}
\end{table*}

In this section, we describe our models, which are summarized in Tab.~\ref{tab:modelsdisk}, where we provide their initial conditions along with some measured quantities.

\subsection{Fiducial run}
\begin{figure}[h!]
\centering
     \includegraphics[width=
          0.49\textwidth]{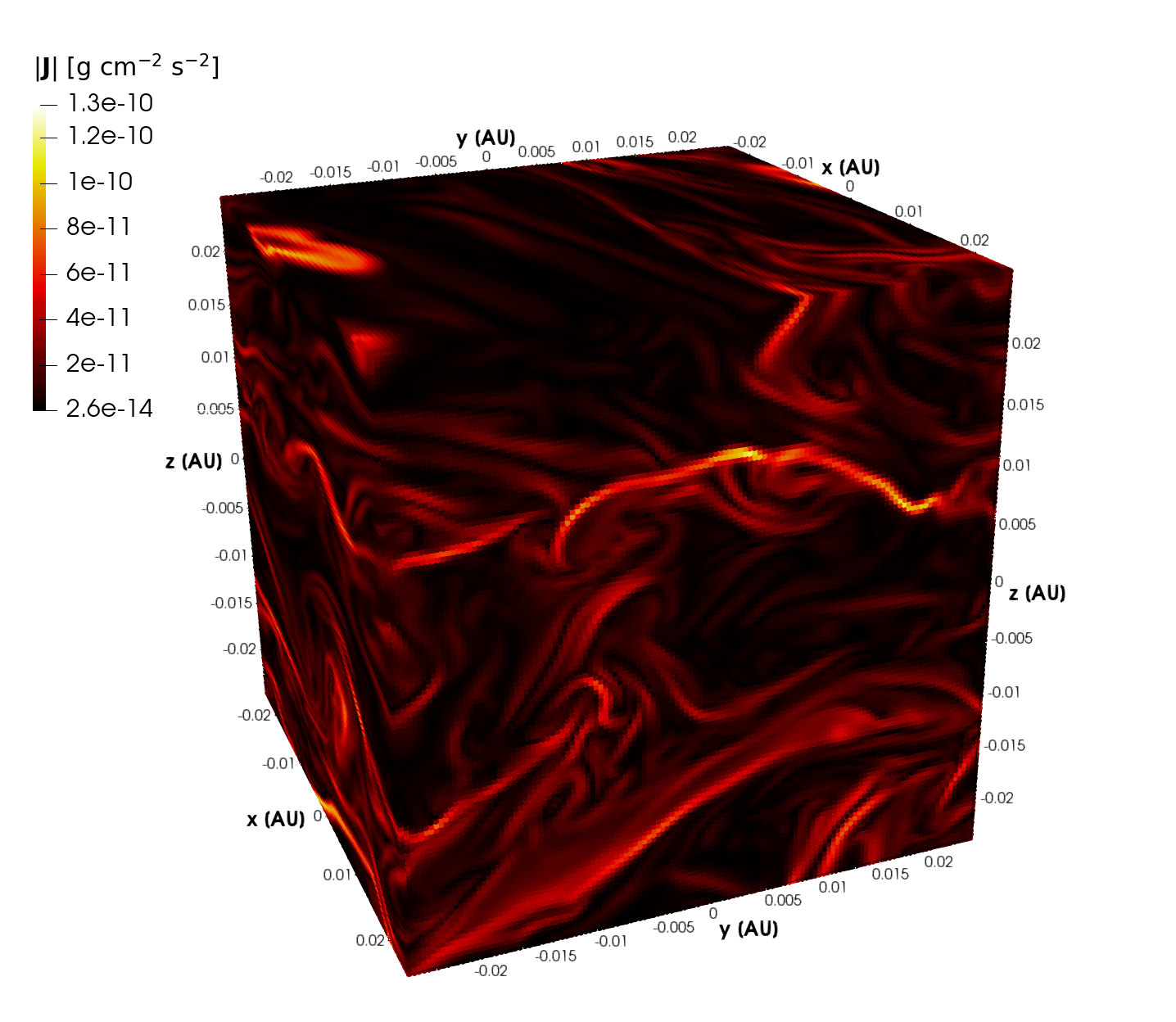}
     \caption{Three-dimensional rendering of the current magnitude for our fiducial model $\textsc{O10A10}$ ($t= 50 \Omega^{-1}$).}
     \label{fig:unstr_fiducial_betaJ}
\end{figure}
\begin{figure*}[h!]
\centering
          \includegraphics[width=
          0.9\textwidth]{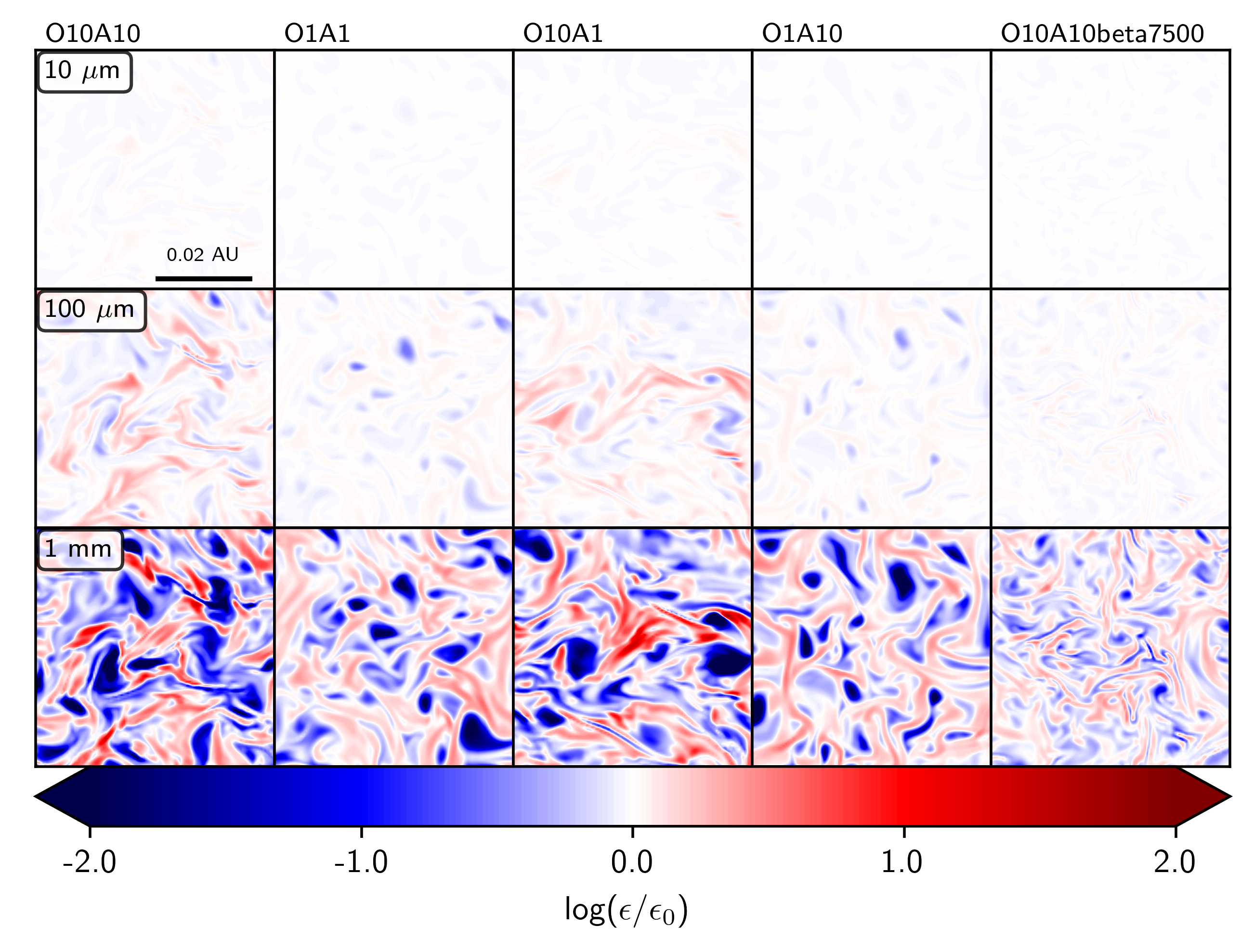}
      \caption{Edge-on slices of the relative dust fraction variations for $\textsc{O10A10}$ (fiducial run), $\textsc{O1A1}$,  $\textsc{O10A1}$ and  $\textsc{O1A10}$ (more resistive runs), from left to right. The slices are displayed at $t\sim 8$ years ($\sim 50 \Omega^{-1}$), except for run $\textsc{O10A10beta7500}$ for which they are displayed at $t\sim 32$ years ($\sim 200 \Omega^{-1}$). The dust grain size increases from top to bottom.}
       \label{fig:unstr_alleps}
\end{figure*}

\begin{figure}[h!]
\centering
          \includegraphics[width=
          0.45\textwidth]{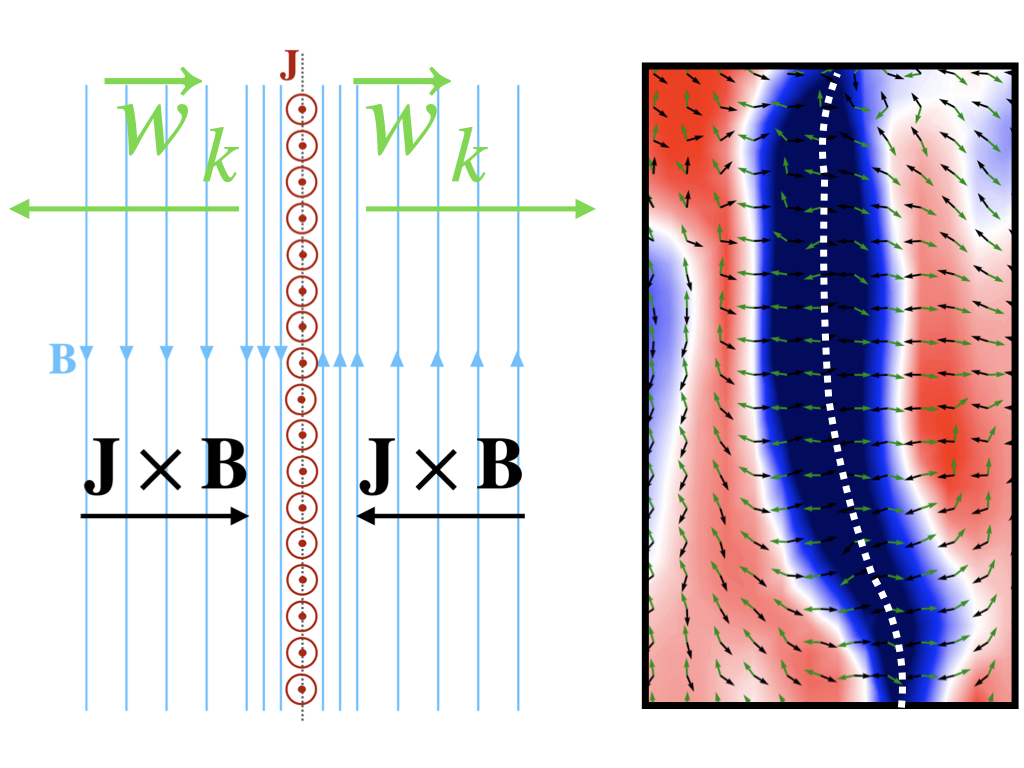}
      \caption{(Left) Schematic local view of a current sheet and the dust drift velocity in its vicinity; $\vec{B}$ is the magnetic field, $\vec{J}$ is the electric current and $\vec{w_k}$ is the dust drift velocity. (Right) Dust ratio variation near a current sheet (zoom in and 90 degree rotation of the left snapshot of Fig. 
      \ref{fig:compnum}; the color scale is the same, with red high and blue low). The dotted line shows the approximate (hand drown) position of the current sheet, the white arrow represent the drift velocity (green) and $\vec{J}\times \vec{B}$ (black) direction projected in the plane of the slice.}
       \label{fig:unstr_cs_illust}
\end{figure}

\begin{figure*}[h!]
\centering

\centering
  \begin{subfigure}[b]{\textwidth}
  \centering
 \includegraphics[width=
          \textwidth]{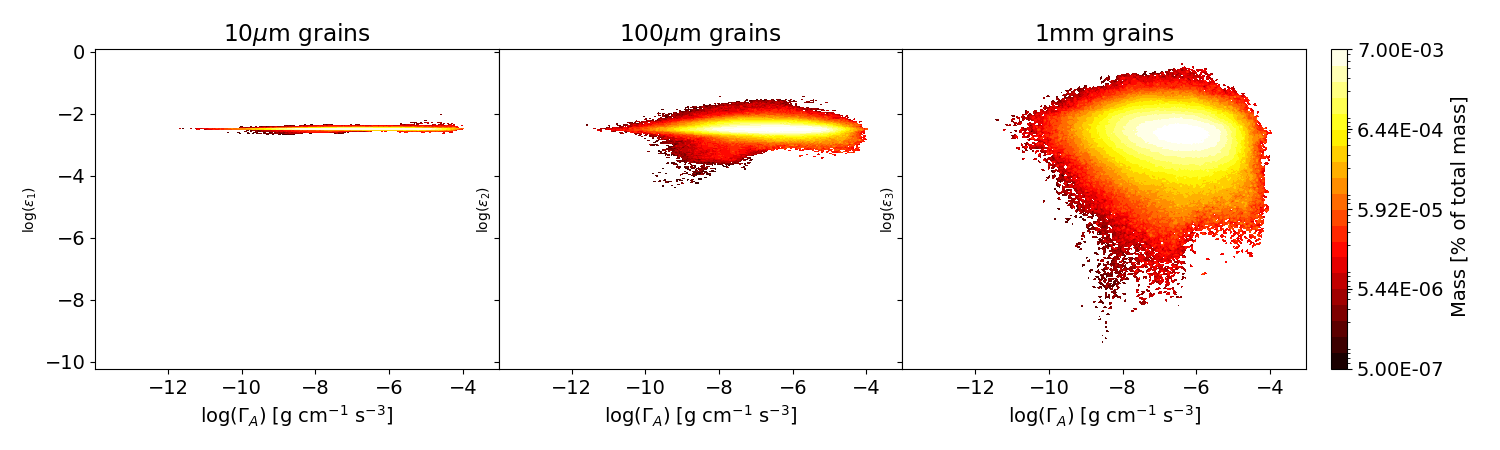}
      \caption{Ambipolar heating source term $\Gamma_{\rm{A}}$. }
       \label{fig:unstr_fiducial_ADeps} 
       
     \end{subfigure}
          
     \begin{subfigure}[b]{\textwidth}
  \centering
 \includegraphics[width=
          \textwidth]{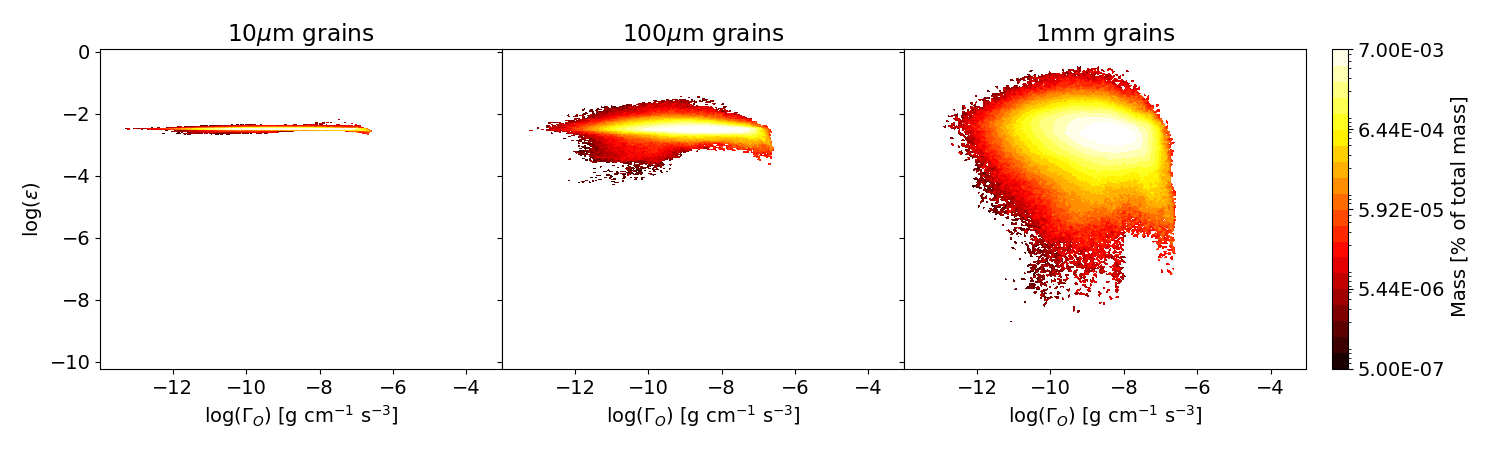}
      \caption{Ohmic heating source term $\Gamma_{\rm{O}}$. }
       \label{fig:unstr_fiducial_omheps} 
       
     \end{subfigure}    
       \caption{Histogram of the dust fraction $\epsilon$ as a function of the ambipolar and Ohmic heating
 source terms for the  $10~\micro\meter$ (left),
$100~\micro\meter$ (middle) and $1~\milli\meter$ grains (right) for
the  $\textsc{O10A10}$ model at $t=50 \Omega^{-1}$. The colors
represent the integrated total mass (the color bar is log-scaled).
Note that the heating is dominated by ambipolar diffusion for all
grain sizes.}
          \label{fig:unstr_fiducial}

\end{figure*}

Our fiducial model, run $\textsc{O10A10}$, is computed at $R=1$~AU, as are all models, and the initial ambipolar and Ohmic Elsasser numbers are both set to $10$. 

Figure \ref{fig:unstr_fiducial_betaJ} shows a three-dimensional rendering of the current magnitude for our fiducial model at the end of the simulation.
As can be seen, due to the moderately low resistivities, quite strong turbulence develops in this model, with $\bar{\alpha}=0.13 $. This value is higher than the value of $\approx 2$--$3 \times 10^{-2} $ from the previous study of \cite{2011ApJ...736..144B}. The difference most likely comes from the use of a 4H x 4 H x 4H box in their case. This is consistent with what we find in Sect.~\ref{sec:numerical} where we explore a larger box and find $\bar{\alpha} \sim 3.7 \times 10^{-2}$. There are strong local variations of $\beta$ ranging from $\sim 0.05$ to $\sim 1000$ and we clearly see sheet-like structures approximately located in the $x$-$y$ plane. These sheets have a typical width of $\sim 10^{-3}~\rm{AU} \approx 1.5 \times 10^{5} ~\kilo\meter$ which is about the same order of magnitude as previous estimates \citep{2004ApJ...606..532J}, but is also roughly $ 4 \Delta x$. We can thus wonder whether the current structures are fully resolved, which we discuss below in Sect.~\ref{sec:numerical}.

In Fig.~\ref{fig:unstr_alleps}, we show slices of the relative dust fraction variations for the $10~\micro\meter$ (top), $100~\micro\meter$ (middle) and $1~\milli\meter$ grains (bottom) for five of the models we computed. The same color scale is used for all grain sizes to best display the range of variation of dust fraction for the $1~\milli\meter$ grains. As can be seen, these grains experience strong dynamical sorting.
Their dust fraction indeed increases up to almost two orders of magnitude in a small fraction of the volume. Smaller $100~\micro\meter$ grains also experience significant, although less important, dust fraction variations of as much as an order of magnitude. The dust fraction variations of $10~\micro\meter$ grains are, however, much smaller (about $\pm 10 \%$ at most).

Dust grains tend to be depleted in current maxima.  This is actually expected, if we consider a plasma with a strong electric current. In this case, the differential dust velocity can be approximated as $\vec{w}_k \approx -t_{\rm{s},k} (\vec{J} \times \vec{B})/\rho$. Generally, dust thus tends to be repelled from the peak of a current sheet where the dust drift velocity reaches a maximum and also flips its direction. This expulsion mechanism is illustrated in Fig.~\ref{fig:unstr_cs_illust}. However, similarly to dust motion in pressure bumps in the case without magnetic field, we expect here the grains to be trapped where $\nabla P_{\mathrm{g}} =\vec{J} \times \vec{B}$. If two current sheets neighbor each other then those traps are necessarily between them.  Note that this concentration mechanism  does not prevent the grains from being completely removed from strongly heated regions because of thermal diffusion and because the ambipolar heating source term $||\vec{J}\times \vec{B}||^2 /||\vec{B}||^2$, which we show below dominates the heating, does not have the same morphology as $||\vec{J}||$.

 In Fig.~\ref{fig:unstr_fiducial}, we display the histogram of the
 dust fraction as a function of the ambipolar and Ohmic heating
 parameters $\Gamma_{\rm{A}}$ and
 $\Gamma_{\rm{O}}$, with the colors representing the integrated
 mass relatively to the total box mass. In the top, right panel, we see that the bulk of the mass of
 the millimeter-sized dust grains resides in regions of moderate
 heating although large dust fractions that seem likely to be
 necessary for chondrule formation are found at a wide range of values
 of  $\Gamma_{\rm{A}}$. As explained earlier, the dust fraction
 variations of small $10~\micro\meter$ grains are much smaller. There is no significant preferential sorting of these
 grains (left panels). As can be seen, the heating source term
 due to ambipolar diffusion dominates
 strongly over the Ohmic source term  $\eta_{\rm{O}}||\vec{J}||^2$ for
 our fiducial model. This is also true for the other models (see
 Tab.~\ref{tab:modelsdisk}). This effect was previously observed by
 \cite{2004ApJ...606..532J}, who noted that the ambipolar heating term
 could exceed the Ohmic one by over an order of magnitude. This could
 have a strong impact on chondrule formation since the Ohmic
 dissipation rates observed by \cite{2004ApJ...606..532J} are similar
 to ours and were already sufficient to produce significant
 temperature variations up to values of $\sim 1500~$K. This however depends on the treatment of the cooling i.e., the values of the opacity, which in turns depend on the abundance and properties of dust grains. 
\subsection{Numerical convergence and box size}

\label{sec:numerical}
\begin{figure*}[h!]
\centering
\begin{subfigure}[a]{0.8\textwidth}
    \includegraphics[width=\textwidth]{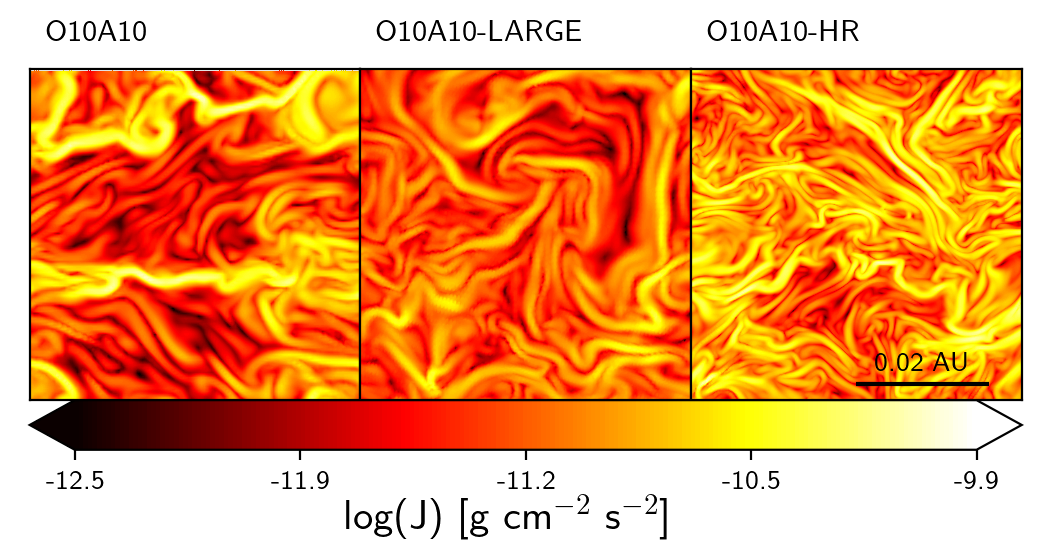}
       \caption{Logarithm of the current norm}
       \label{fig:comp_eps} 
     \end{subfigure}
     \begin{subfigure}[b]{0.8\textwidth}
    \includegraphics[width=\textwidth]{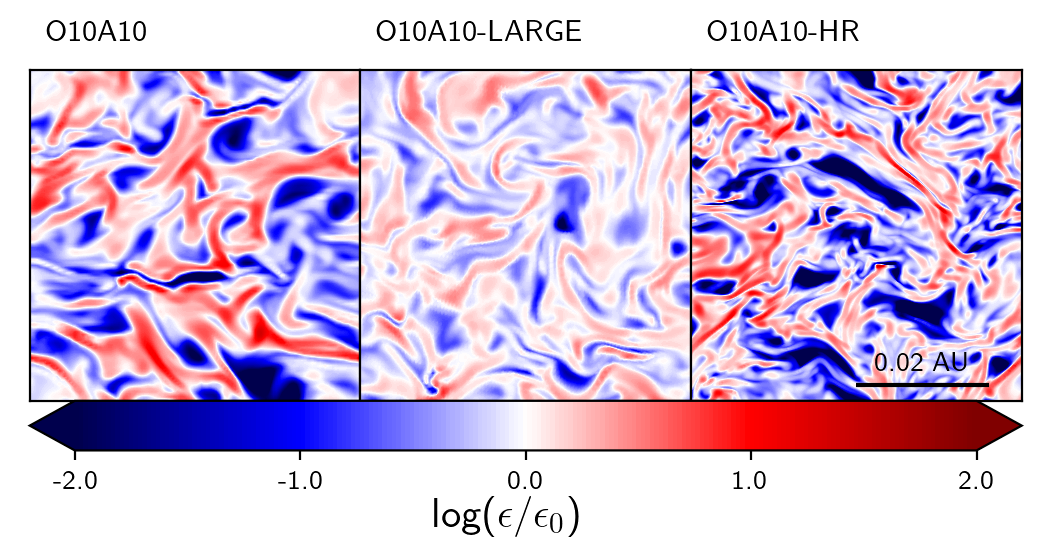}
       \caption{Dust fraction variation of millimeter grains}
       \label{fig:comp_J} 
     \end{subfigure}
     \caption{
       Slices in the $x$-$z$ plane of the fiducial model
         $\textsc{O10A10}$ (left), the large box model
         $\textsc{O10A10-LARGE}$ (center), and the high-resolution
         model $\textsc{O10A10-HR}$ (right). All models are
         displayed at the same size, so we have zoomed out by a factor of
         two in the case of $\textsc{O10A10-LARGE}$, which has a
         box size of 0.1~AU, compared to 0.05~AU for the other two models. 
       \label{fig:compnum} }
\end{figure*}

We now discuss the impact of the box size and resolution by
introducing two additional models, runs  $\textsc{O10A10-Large}$ and
$\textsc{O10A10-HR}$. The first one, run  $\textsc{O10A10-Large}$, is
the same as the fiducial model, run $\textsc{O10A10}$, but with a twice
as large box, using a $256^3$ grid to maintain constant resolution. The
second model,  $\textsc{O10A10-HR}$, is computed with the same box
size as the fiducial model, but using a $256^3$ grid to double the
numerical resolution. As $\textsc{O10A10-HR}$ is computationally
expensive because of the quadratically reduced ambipolar diffusion timestep, we compare the three models at $t=20 \Omega^{-1}$. 

Figures \ref{fig:comp_eps} and \ref{fig:comp_J} show that, although the three models
are qualitatively similar, they show some notable differences. First, we see that increasing the box size seems to lead to a more homogeneous current sheet distribution. This
suggests that the two dominant current sheets observed in the
fiducial model might be of numerical origin.  As noted in previous
studies of MRI turbulence in shearing box simulations, increasing the
box size leads to less efficient turbulent transport, which is also
why the dust fraction variations are smaller in
run $\textsc{O10A10-LARGE}$ (the dust fraction increases by a factor up to 18, against $\sim56$ for the fiducial run and $\sim51$ for $\textsc{O10A10-HR}$). This is in line with the fact that $\bar{\alpha}\sim 3.7 \times 10^{-2}$ for \textsc{O10A10-LARGE}, which is about 3 times smaller than for the fiducial run.  This particular and essential detail
encourages future calculations that should consider current sheet
formation but in a larger scale environment. This could be done either with stratified
shearing boxes, as we plan to do in future investigations, or in global
calculations, for which the achievable resolution remains largely insufficient.

In terms of resolution, a comparison of run $\textsc{O10A10-HR}$ with
our fiducial one seems to indicate that we are approaching convergence
both in terms of thickness of the current sheets, which is not a
factor of two smaller in run $\textsc{O10A10-HR}$ at twice the
resolution, and also in terms of the range of variations for the current norm (the peak current at $t=20 \Omega^{-1}$ is $1.5 \times 10^{-10}~\mathrm{g} ~\mathrm{cm}^{-2}~\mathrm{s}^{-2} $ for $\textsc{O10A10-HR}$, compared to $1.4 \times 10^{-10} ~\mathrm{g}~\mathrm{cm}^{-2}~\mathrm{s}^{-2}$ in the fiducial run) and dust fraction (the dust ratio increases up to a factor of $\sim50$--60 for $\textsc{O10A10-HR}$ and $\textsc{O10A10}$). However, in terms of spatial distribution of the
current sheets, increasing the resolution seems to have a similar effect as increasing the box size. As can be seen, current sheets are more evenly distributed in model $\textsc{O10A10-HR}$ than they are in the fiducial model. We show in Sect.~\ref{sec:pdfs} that this only has a little effect on the probability distribution functions of the dust in the current sheets.

\subsection{Impact of the resistivity}

\begin{figure}[h!]
\centering
     \includegraphics[width=
          0.49\textwidth]{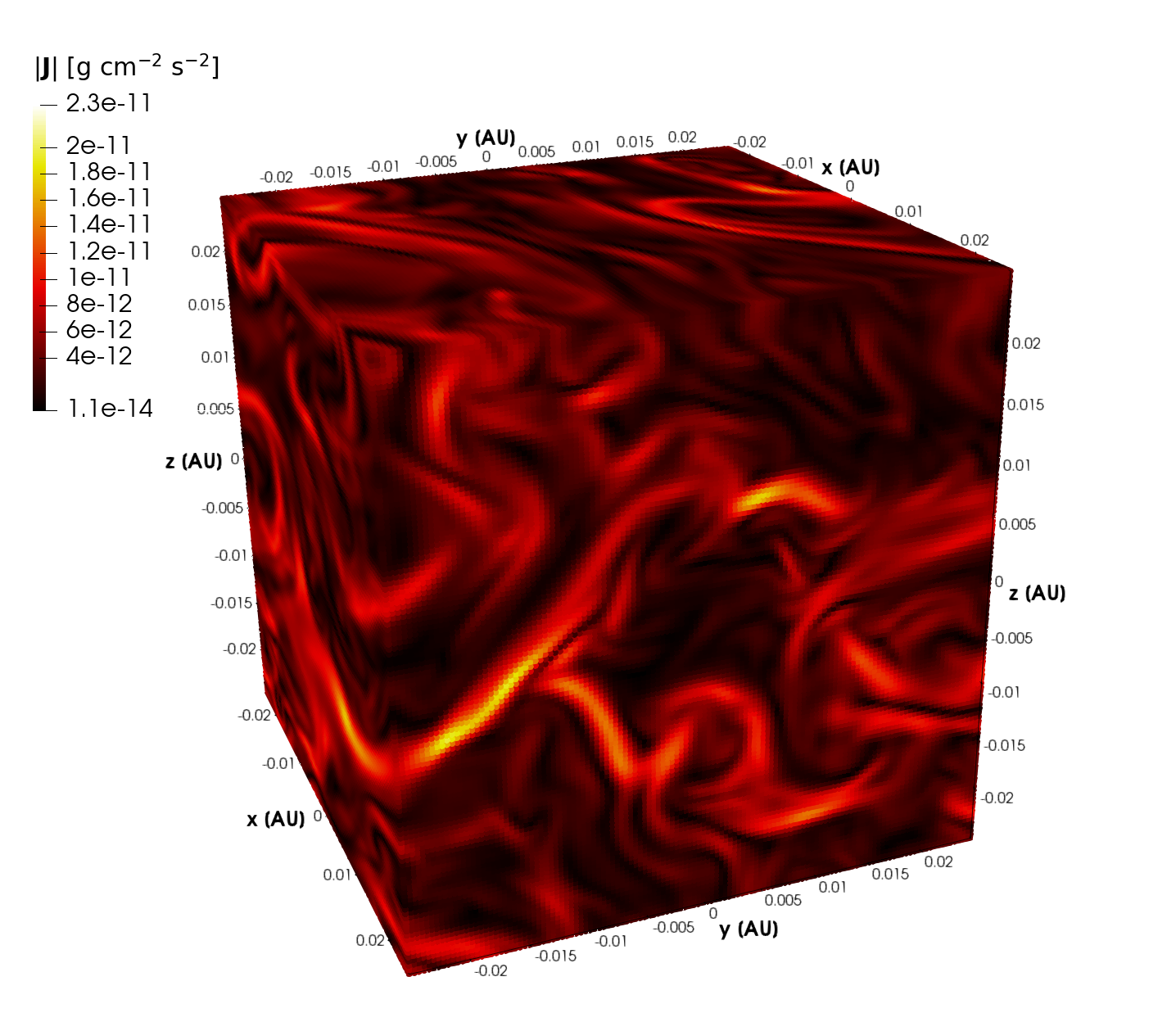}
     \hfill
     \caption{Three-dimensional rendering of the current magnitude for our most resistive model $\textsc{O1A1}$ ($t= 50 \Omega^{-1}$).}
     \label{fig:unstr_res_betaJ} 
\end{figure}

We next explore the impact of the value of the resistivities with model $\textsc{O1A1}$, where the two Elsasser numbers are set to unity (and hence the model is more resistive than our fiducial one). 
Fig.~\ref{fig:unstr_res_betaJ}  shows that this model forms current sheet structures as in the fiducial run. In the case of run $\textsc{O1A1}$ however, these sheets are wider than for the fiducial case with a typical thickness of $\sim 2 \times 10^{-3}~\rm{AU}$. A thickening of the current structures is expected with increasing resistivity. Indeed, as the ambipolar length increases, the magnetic field lines are rearranged over larger scales. As expected, the MRI turbulence is weaker in $\textsc{O1A1}$ than in the fiducial case. We measure $\bar{\alpha}=2.5 \times 10^{-2}$, which is about one order of magnitude smaller than the fiducial value. 
 
Local variations of the dust fraction for the 100 micron- and millimeter-size grains are similar to those of run $\textsc{O10A10}$ although
not as strong can be seen in Fig.~\ref{fig:unstr_alleps}. Moreover, the regions of high dust fraction are thicker in this model because the width of the current sheets is larger than in model $\textsc{O10A10}$. In this model, the dust still tends to be expelled from regions of maximal current. However, as explained earlier this does not necessarily mean that dust is expelled from regions of high dissipation by ambipolar diffusion. As in the fiducial model, small $10~\micro\meter$ grains remain well coupled to the gas everywhere in the box and their variation of concentration is insignificant. 

We also computed two additional models  $\textsc{O10A1}$ and $\textsc{O1A10}$, where the two Elsasser number are different ($\Lambda=10$ and $\mathrm{Am}=1$ for $\textsc{O10A1}$ and the opposite for $\textsc{O1A10}$). We still form current sheets which, as expected, are intermediate in size between $\textsc{O10A10}$ and $\textsc{O1A1}$. For the same reasons, the dust concentration variations are also stronger in these two additional models than in $\textsc{O1A1}$ and less important than in $\textsc{O10A10}$. For the same reason as for $\textsc{O1A1}$, the turbulence is also weaker for $\textsc{O1A10}$ and  $\textsc{O10A1}$.

\subsection{Impact of the plasma $\beta$}

\begin{figure}[h!]
\centering
     \includegraphics[width=
          0.49\textwidth]{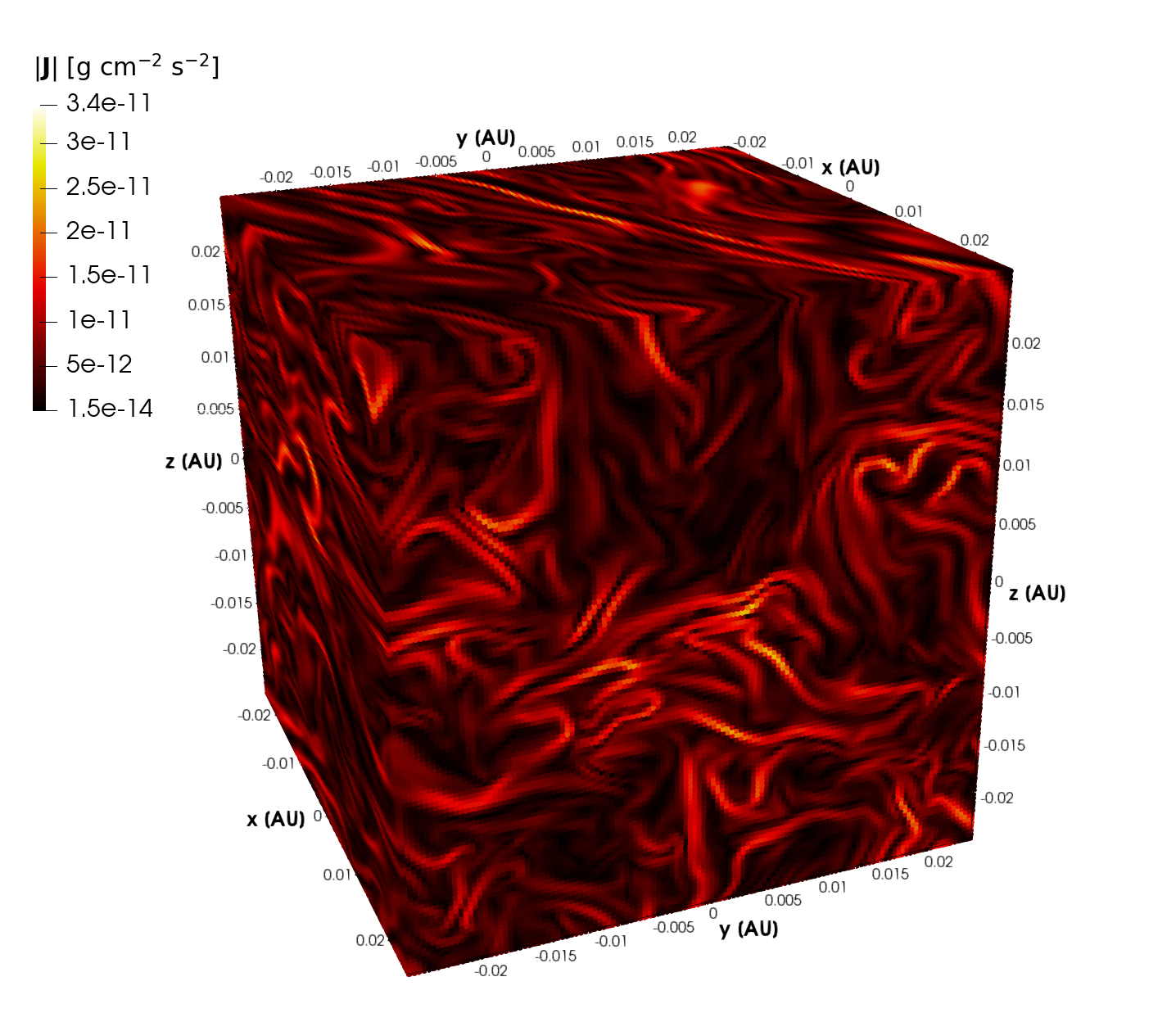}
     \hfill
     \caption{Three-dimensional rendering of the current magnitude for high-$\beta$ model $\textsc{O10A10beta7500}$ ($t= 80 \Omega^{-1}$).}
     \label{fig:unstr_fid_beta10} 
\end{figure}

We finally investigate the effect of the initial magnetization with the
$\textsc{O10A10beta7500}$ model that is the same as our fiducial
run, but with a weaker field so that $\beta =7500$. 

Figure~\ref{fig:unstr_fid_beta10} shows the current sheets of this model
at $t=200~\Omega^{-1}$, as it takes a longer time to reach a steady
state because the growth rate of the MRI is lower. In this model $\bar{\alpha}=3 \times 10^{-3}$, a value lower than the fiducial. This is expected from a run with a higher initial $\beta$ \citep[see Fig. 4 of][]{2011ApJ...736..144B}. As we can see, this model still forms a large number of current
sheets with a typical width that is still of the order of
$10^{-3}~$AU. In this model, the peak of the current is about three
times lower than in the case of $\textsc{O10A10}$. However, on
average, the two models are comparable in terms of current magnitude, so we can
expect the conditions in $\textsc{O10A10beta7500}$ to produce
significant local changes of temperature as well. We also note that in this
model the current sheets are more uniformly distributed than in the
fiducial model. This is probably an effect of the box size being a larger multiple of $\lambda_{\rm{c}}$ (see the discussion in Sect.~\ref{sec:numerical}).

Figure~\ref{fig:unstr_alleps} shows that the millimeter dust grains
experience significant variations of concentration in this
model. These variations are comparable to the ones observed in the
high resistivity model $\textsc{O1A1}$, but are less important
than for the fiducial model, since the turbulence is weaker. For the
same reasons and again similarly to the fiducial model, grains of size less than $100 \micro\meter$ do not experience significant dust fraction variations in this model.

\section{Discussion and summary}
\label{sec:discuss}

\subsection{Dust distribution in current sheets}
\label{sec:pdfs}

\begin{figure*}[h!]
\centering
  \begin{subfigure}[a]{0.8\textwidth}
    \includegraphics[width=\textwidth]{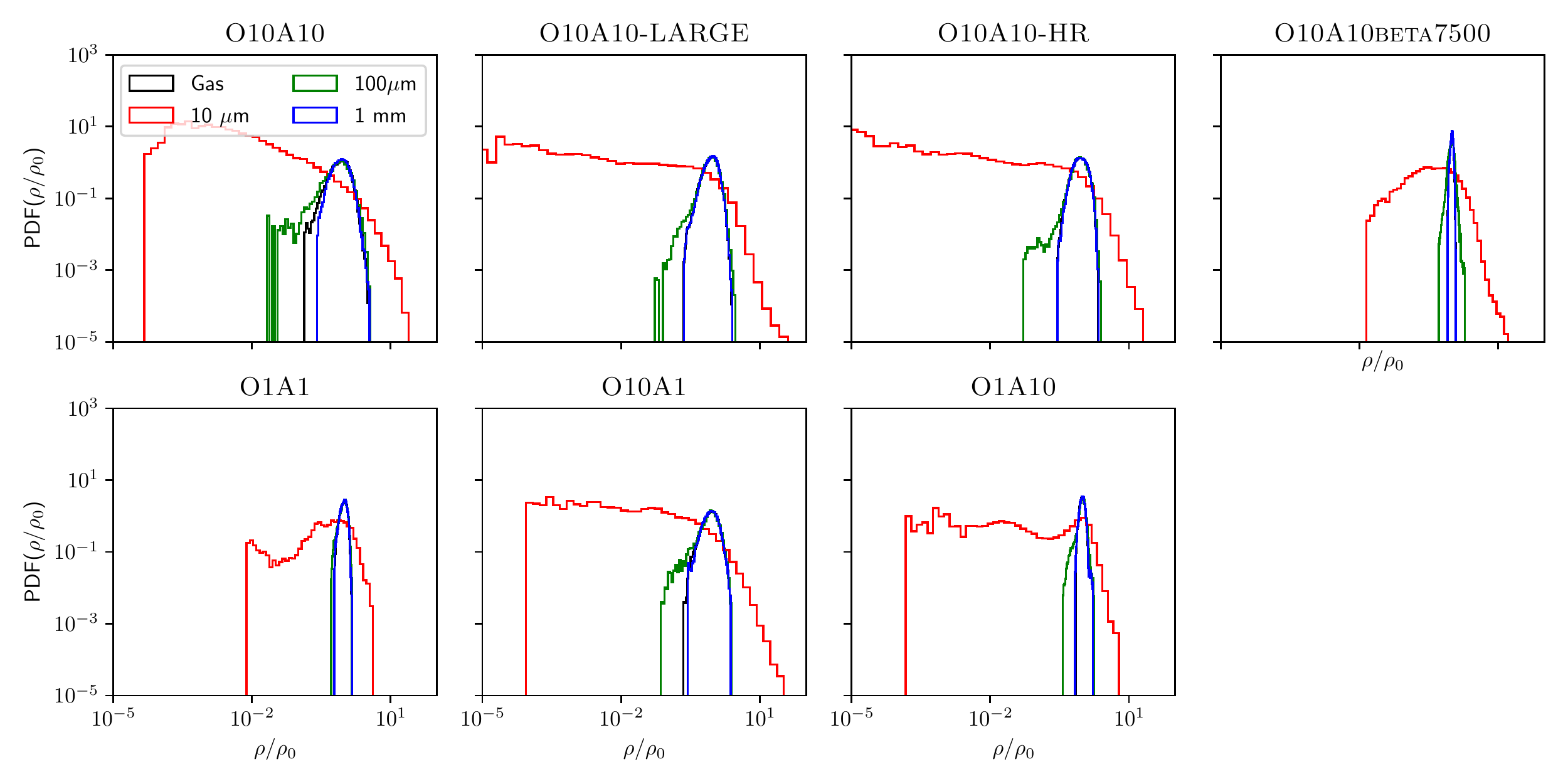}
       \caption{Strong ambipolar diffusion heating regions.}
       \label{fig:pdfsa} 
     \end{subfigure}
       \begin{subfigure}[b]{0.8\textwidth}
    \includegraphics[width=\textwidth]{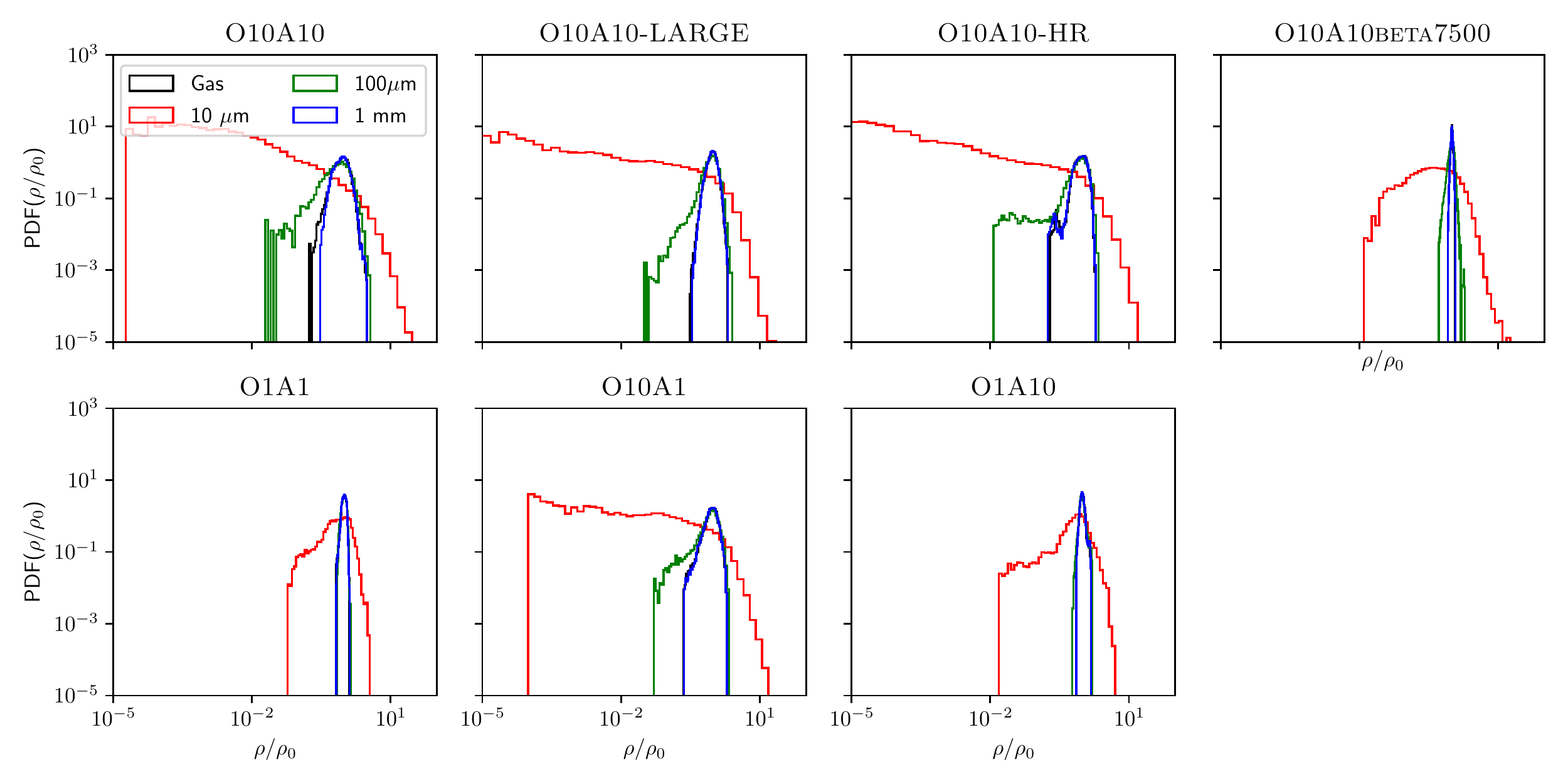}
       \caption{Strong Ohmic resistivity heating regions.}
       \label{fig:pdfsb} 
     \end{subfigure}
       \begin{subfigure}[c]{0.8\textwidth}
    \includegraphics[width=\textwidth]{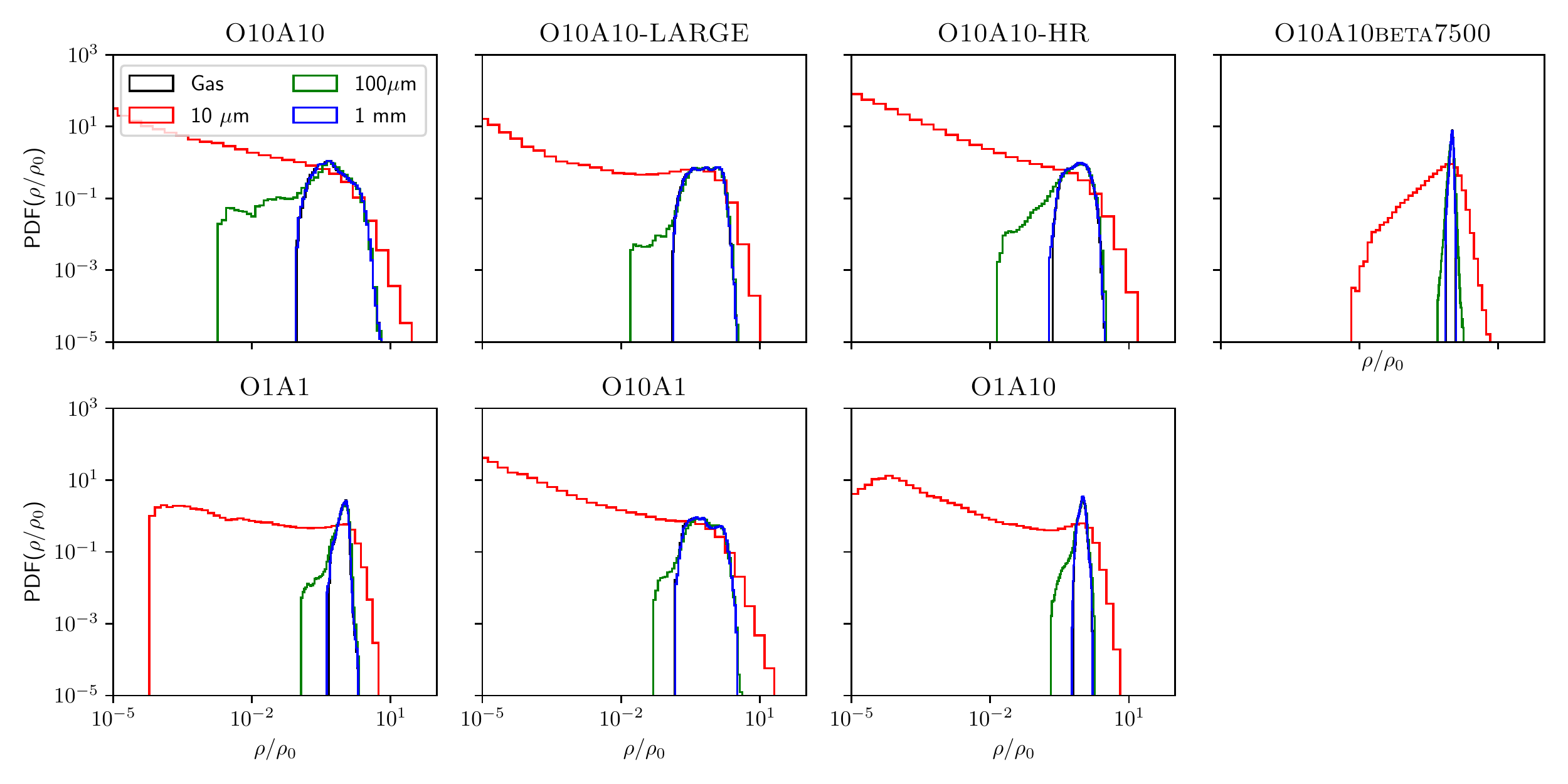}
       \caption{Whole box.}
       \label{fig:pdfsc} 
     \end{subfigure}
     \caption{Density PDFs of the three dust sizes and the gas,
       scaled by the initial values of the density. 
     \label{fig:pdfs} }
\end{figure*}

In this section, we discuss the impact of the magnetic turbulence and
the intensity of ambipolar diffusion and Ohmic dissipation on the dust distribution
in the current sheets and in the whole box in our models. We show, in
Fig.~\ref{fig:pdfs}, the  probability density
 functions (PDFs) of the three
dust sizes and the normalised gas density. Fig.~\ref{fig:pdfsa}
shows the PDF in
ambipolar current sheets (defined in Eq.~\ref{eq:defcs}, thresholds given in Tab.~\ref{tab:modelsdisk}),
Fig.~\ref{fig:pdfsb} shows the PDFs in Ohmic current sheets (defined similarly), and finally Fig.~\ref{fig:pdfsc} shows them for the whole box.
For all the models millimeter-sized grains
experience strong dynamical sorting with respect to the gas, leading
to high dust densities and concentrations (since the gas PDFs have a
narrower width). We also quite clearly see that the models developing the stronger turbulence are the ones showing the strongest dust sorting, i.e O10A10 (for the two resolutions and box size) and O10A1. Nevertheless, all models show at least an order of magnitude increase of the maximum dust-to-gas ratio for mm-sized grains.

Let us now compare the different models. The PDFs from
$\textsc{O10A10}$, $\textsc{O10A10-HR}$ and
$\textsc{O10A10-LARGE}$ show strong similarities, which indicate
that even though $\textsc{O10A10}$ might not resolve the current
sheets very well, it captures the most important results of
this study: the formation of regions of strong dissipation and the
strong dynamical sorting of millimeter-sized grains. Not only the dust but also the gas PDF both have a narrower range in the more resistive runs because the ambipolar diffusion and the Ohmic resistivity
significantly damp the MRI for this model. Interestingly, and although it does not generate strong gas density variations, the turbulence generated in the $\textsc{O10A10beta7500}$ model actually also leads to an extended high density tail for the millimeter-sized grains quite similar to the one observed in our fiducial model. This is because the cause of the sorting is actually the strong current and not the pressure gradients.

In short, significant local dust fraction variations leading to
very localized regions of high concentration and dust densities
typically occur in our models. We also show that the amplitude of these variation is controlled by the strength of the MRI turbulence.

\subsection{Implications for chondrule formation}

Noticeably, the high density tail of the PDF of millimeter-sized
grains, although more extended in the case of the whole box, is relatively similar in all regions, i.e. whether the current is strong or not.  This is interesting if we assume that chondrules do indeed form in flashes (although we recall that many other theories exist). Indeed, strong dust concentration is a prerequisite for chondrule formation, to explain the abundance of volatile elements
in chondrules \citep[][]{2008Sci...320.1617A}. We see in Fig.~\ref{fig:pdfs} that it is as likely to happen in regions of strong dissipation as
anywhere else in the box. Nevertheless, strong heating events at high
dust concentration (i.e. flashes) involve only a small fraction of the dust mass and of the box
volume, as the PDF tail decreases sharply with density.  We indeed find that current sheets represent about $1.5\%$ of the box volume in the fiducial case ($1.78\%$ in the $\textsc{O10A10-LARGE}$ run and $1.2\%$ in run $\textsc{O10A10-HR}$) according to the definition of  Eq.~\ref{eq:defcs}.  In all the models, they represent about $1-2\%$ of the total volume. The condition of the rarity of flash occurrences, which  is also a
prerequisite for a successful chondrule formation theory is therefore met here. In addition,
small grains that are the precursors of the matrix are made from the
same material as chondrules but should not experience flashes as often
as larger grains in order to survive in the disk, as can be seen from their uniform distribution with a PDF
very similar to that of the gas. 

We point out that the parameter space for current sheets that could be suitable for chondrule formation is likely to be narrow. For smaller Elsasser numbers (as in the dead zones) than presented here, the MRI is completely damped and therefore no current sheets can form. For larger values, we expect current sheets to form (as they would already form in ideal MHD) but the dissipation rates will also decrease (and eventually become negligible) because they are proportional to the resistivities. Nevertheless, since the inner radii and upper regions of protoplanetary disks are strongly ionized, the Elsasser numbers must always pass through unity at the edge of the dead zone. Arguably, to have a narrow parameter space for strongly dissipative current sheets is in fact convenient to explain chondrule formation since flashes must remain localised events.   

\subsection{Caveats}

We now discuss the various caveats of this study. First of all, as current sheets are very narrow structures, we have to rely on small scale unstratified simulations in order to attempt resolving them. Although we seem to be reaching convergence for our $256^3$ model, we noted an effect of the box size on the strength of the turbulence in our fiducial model. In addition, it is not yet clear that these current sheets would form in a stratified disk. Stratified resistive disks often have low midplane Elsasser numbers that produce disk winds rather than triggering the MRI \citep{2013ApJ...769...76B,2014A&A...566A..56L}, with only very little turbulence in the disk. Some studies \citep[e.g.,][]{2015ApJ...801...84G} did find current sheets  at the edge of the dead zone in their stratified models, although they were not MRI active and were not as resolved as in our models. It is also worth pointing out that a midplane current sheet is also typically reported when disk winds are present \citep[e.g.,][]{BaiStone2017}. Interestingly \cite{2021A&A...650A..35L} have shown that Ohmic dissipation could push it to the layers of the disk. High resolution studies of stratified disks will need to be computed in order to confirm that current sheet formation happens systematically at the edge of the dead zone.  

The long term goal of our study is to better understand chondrule formation. Again, stratification could prove to be an important aspect in the matter. The initial conditions that we explored
are designed to reproduce quite well the inner regions of
protoplanetary disks (at $R=1~AU$ and $z\approx 1$--$2H$), where
chondrule formation in current sheets is expected to happen. As they reproduce conditions above the midplane, the chondrule precursors would need to be lifted efficiently so that they can be reprocessed by these current sheets. Adding
stratification in future calculations will allow assessing whether sufficiently
large numbers of millimeter- and 100 micron-sized grains can be lifted high enough to form the observed chondrule population. If the amount of lifted dust material is insufficient, the observed variations of dust fractions might not be enough to generate the high dust concentrations that are also required. Stratification could also give birth to new interesting structures such as rings \citep[e.g.][]{Bethune2017} or lead to vertical shear instability if the MRI is completely damped or strongly saturated by ambipolar diffusion \citep{Latter2022}. These structures/instabilities would affect the dust dynamics and might be key for the transport of chondrules precursors.

In addition, the models that we presented here are isothermal for simplicity and because we focus on the dust dynamics. We can still measure the heating rate by Ohmic dissipation and ambipolar diffusion and have shown that they are comparable to \cite{2014ApJ...791...62M} who observed strong temperature variation. However depending on the cooling rate, that relies on the choice of dust distribution and opacity model, this heating might still be insufficient to form chondrules. This justifies further exploration accounting for the thermal evolution of the disk.

Finally, aside from the $B^2$ dependency of the ambipolar resistivity, we have imposed constant resistivities in our models. This approach is justified because it allows for a simple parameterisation of the Ohmic and ambipolar diffusion with the dimensionless Elsasser numbers without requiring any chemical network. However, the downside of the approach is that the dependency of the resistivity with the density and temperature are not taken into account. As was demonstrated by \cite{2012ApJ...761...58H}, short-circuit instabilities could narrow current sheets significantly leading to very high temperatures provided that the resistivity decreases with an increasing temperature. \cite{2015ApJ...811..156D} later have argued that these conditions are likely not met with realistic diffusion rates of volatile alkali metals out of dust grains. However, using a constant resistivity, \cite{2014ApJ...791...62M} still reported temperature increases sufficient for chondrule formation in current sheets. The latter instability might thus might not be necessary to explain chondrule formation in current sheets. Future high resolution models should assess current sheet formation/evolution with more advanced chemical networks.

\subsection{Summary and prospects}
\label{sec:conclu}
 In this article, we investigated the dynamics of dust grains with
 properties similar to chondrules, within unstratified shearing box
 \texttt{RAMSES} calculations that aim to resolve the formation of
 dissipative current sheets through the non-ideal MHD effects of ambipolar diffusion and Ohmic resistivity.
 
 We investigated the effect of the strength of the Ohmic resistivity
 and ambipolar diffusion, the initial density, and magnetic field
 strength, as well as the numerical resolution and the box size. Our main findings are:
 \begin{itemize}
     \item Current sheets form with typical widths of 1--2$\times 10^{-3}$~AU and strong dissipation rates \citep[as in][]{2014ApJ...791...62M}.
      \item Ambipolar diffusion systematically produces dissipation
        rates more than an order of magnitude higher than those produced by Ohmic
        resistivity.
   \item These current sheets could produce intermittent, high temperature hot spots
        in the regions of protoplanetary disks  if the cooling rates are not too fast where this effect dominates. 
    \item  We observe dust fraction variations of up to almost two
      orders of magnitude for all initial conditions that we studied.
      These variations are directly connected with current sheet
      formation, as the dust is typically repelled from the peaks of
      the current sheets but concentrates in their envelopes.
    \item The regions of strong millimeter-sized grain concentration are
      highly localized,
      as required in a successful chondrule formation theory
      \citep[e.\ g., ][]{2008Sci...320.1617A}.
 \end{itemize}

 All the models described here were computed in the isothermal
 approximation,  so we cannot directly derive the temperatures or cooling rates of the
 hot spots, as would be needed to study chondrule formation. In future
 work, we plan to include the effect of the ambipolar and Ohmic heating
 source terms in non-isothermal models. As our initial conditions
   are optically thick, those models must include radiative
   transfer. This will be computed in the flux-limited diffusion
 approximation using the solver of \cite{2011A&A...529A..35C}
to model diffusion from heated regions.
 Those models will also include a global cooling at the orbital timescale. They will allow us to assess the formation of hot spots in unstratified shearing box calculations. Inclusion of stratification will allow study of current sheet structures 
that could form even in the absence of MRI, as seen for example in \citet{2015ApJ...801...84G}.

 \begin{acknowledgements}
 We thank the referee for providing useful comments that helped to improve our manuscript.
We acknowledge financial support from the Programme
National de Physique Stellaire (PNPS) of CNRS/INSU, CEA, and CNES,
France. This work was granted access to the HPC resources of CINES (Occigen) and IDRIS (Jean Zay) under the allocation DARI A0080407247 made by GENCI. Computations
were also performed at the Common Computing Facility (CCF) of the LABEX
Lyon Institute of Origins (ANR-10-LABX-0066).
Ugo Lebreuilly acknowledges financial support from the Annette Kade
fellowship of the AMNH, from the Ecole Normale Supérieure Paris-Saclay via the Contrats doctoraux spécifiques pour normaliens (CDSN) and from the European Research Council (ERC) via the ERC Synergy Grant ECOGAL (grant 855130). This project was partly supported by the IDEXLyon project (contract n ANR-16-IDEX-0005) under University of Lyon auspices. The plots were generated using the OSYRIS library developed by Neil Vaytet.
We thank Prof. S\'ebastien Fromang for very useful discussions and comments on our manuscript. 
\end{acknowledgements}
\bibliographystyle{aa}
\bibliography{ref}

\end{document}